\documentclass[twoside]{dis04}

\def\runauthor{Fridger Schrempp and Andre Utermann}
\def\shorttitle{QCD Instanton-driven Gluon Saturation}
\usepackage{amsmath,amssymb}
\usepackage{bbm}
\usepackage{cite}
\newcommand{\lwig}{\mbox{\,\raisebox{.3ex}
    {$<$}$\!\!\!\!\!$\raisebox{-.9ex}{$\sim$}\,}}
\newcommand{\gwig}{\mbox{\,\raisebox{.3ex}
    {$>$}$\!\!\!\!\!$\raisebox{-.9ex}{$\sim$}}\,}
\newcommand{\rav}{\langle\rho\rangle}
\newcommand{\xbj}{x_{\rm Bj}}

\newcommand{\vl}{\mathbf{l}}
\newcommand{\vr}{\mathbf{r}}

\newcommand{\sidp}{\sigma_\text{\tiny DP}}

\def\Journal#1#2#3#4{{#1}{\bf #2}, #3 (#4)}
\def\NPA{{\em Nucl. Phys.} {\bf A}}
\def\NPB{{\em Nucl. Phys.} {\bf B}}
\def\PLB{{\em Phys. Lett.}  {\bf B}}

\def\PRD{{\em Phys. Rev.} {\bf D}}

\def\ZPC{{\em Z. Phys.} {\bf C}}

\def\CPC{\em Comput. Phys. Commun. }

\def\JPG{{\em J. Phys.} {\bf G}}

\def\APPB{{\em Acta Phys.\ Polon.} {\bf B}}
 
\def\PRP{\em Phys. Rept. }
\begin{document}
\title{Instanton-driven Gluon Saturation}

\author{F. SCHREMPP} 

\address{Deutsches Elektronen Synchrotron (DESY), \\
Notkestrasse 85, \\ 
D-22607 Hamburg, Germany\\ 
E-mail: fridger.schrempp@desy.de}

\author{A. UTERMANN}
\address{Institut  f{\"u}r Theoretische Physik der
Universit{\"a}t Heidelberg,\\
Philosophenweg 16, \\
D-69120 Heidelberg, Germany\\
E-mail: A.Utermann@thphys.uni-heidelberg.de
}

\maketitle

\abstracts{We report on the interesting possibility of
instanton-driven gluon saturation in lepton-nucleon scattering at
small Bjorken-$x$.   
The explicitly known instanton gauge field serves as a concrete
realization of an underlying non-perturbative saturation mechanism
associated with strong classical fields. The conspicuous, intrinsic
instanton size scale known from lattice simulations, turns out to
determine the saturation scale. The ``colour 
glass condensate'' can be identified in our approach with the
QCD-sphaleron state, dominating instanton-induced processes in 
the softer regime.}

Lepton-nucleon scattering at small Bjorken-$x$ uncovers a novel regime
of QCD, where the coupling 
$\alpha_{\rm s}$ is (still) small, but the parton densities are so large
that conventional perturbation theory ceases to be applicable. 
In general, one expects non-linear corrections to the evolution equations
\cite{gribovnu} to arise and to become
significant in this regime, potentially taming the growth of the gluon
distribution towards a ``saturating'' behaviour.
Much interest has recently been generated through association of the
saturation phenomenon with a multiparticle quantum state of high
occupation numbers, the ``Colour Glass Condensate'' that  correspondingly,
can be viewed~\cite{cgc} as a strong {\em classical} colour field
$\propto 1/\sqrt{\alpha_{\rm s}}$. 

In this contribution, we shall briefly summarize our results on the interesting
possibility of instanton-driven saturation at small Bjorken-$x$. 
Being extended non-perturbative and topologically non-trivial
fluctuations of the gluon field, instantons~\cite{bpst} ($I$) represent
a fundamental {\em non-perturbative} aspect of
QCD. Notably, the functional form of the instanton gauge 
field $A_\mu^{(I)}$ is explicitely known and its strength is $\propto
1/\sqrt{\alpha_{\rm s}}$, just as needed. In addition, we shall summarize
why an identification of the ``Colour Glass Condensate'' with the
QCD-sphaleron state appears very suggestive \cite{su2,su3}.  From
$I$-perturbation theory we learned, that the instanton contribution tends
to strongly increase towards the softer
regime~\cite{rs1,mrs,rs2,qcdins}, were saturation effects are expected
to occur. Our results crucially rely on non-perturbative information
from high-quality lattice simulations~\cite{ukqcd,rs-lat,rs3}. For
related approaches associating instantons with high-energy scattering, see
Refs.~\mbox{\cite{levin,shuryak1}}. Instantons in the
context of small-$x$ saturation have also been studied recently by
Shuryak and Zahed~\cite{shuryak3}, with conclusions differing in part 
from those of our preceeding work~\cite{fs,su1,su2,su3}. Their main
emphasis rests on Wilson loop scattering, and lattice information was
not used in their approach.

An investigation of saturation in $\gamma^\ast P$ scattering becomes
most transparent in the 
colour-dipole picture~\cite{dipole}. At high energies, 
the lifetime  of the $q\overline{q}$-dipole, into which the incoming
$\gamma^\ast$ fluctuates, is much longer than the
interaction time between this $q\overline{q}$-pair and the hadron.
For small $\xbj$, this gives  rise to the familiar factorized
expression of the inclusive photon-proton cross sections, 
\vspace{-0.2cm}
\begin{equation}
\sigma_{L,T}(\xbj,Q^2) 
=\int_0^1 d z \int d^2\vr\; |\Psi_{L,T}(z,r)|^2\,\sigma_{\mbox{\tiny DP}}(r,\ldots)\,.  
\label{dipole-cross}
\end{equation}
Here, $|\Psi_{L,T}(z,r)|^2$ denotes the modulus squared of the 
(light-cone) wave function of the virtual photon, calculable in pQCD,
and $\sigma_{\mbox{\tiny DP}}(r,\ldots)$ is the
$q\overline{q}$-dipole\,-\,nucleon cross section.  The variables are
the transverse $(q\overline{q})$-size $\mathbf r $ 
and the photon's longitudinal momentum fraction $z$ carried by the quark. 
The dipole cross section includes the main
non-perturbative contributions. Within
pQCD~\cite{dipole,dipole-pqcd}, $\sigma_{\mbox{\tiny DP}}$ is known to
vanish with the area $\pi r^2$ of the $q\overline{q}$-dipole. Besides
this phenomenon of ``colour transparency'' for small $r=|\vr|$,  the
dipole cross section is expected to saturate towards a constant, once the
$q\overline{q}$-separation $r$ exceeds a certain saturation scale $r_s$. 

In our study, the question is: Can background instantons of size
$\sim\langle\rho\rangle$ give rise to a saturating form
 $\sidp^{(I)}(r,\ldots)\propto
\langle\rho\rangle^2$ for $r\gwig \langle\rho\rangle$? Our 
strategy was to start from $I$-perturbation theory~\cite{mrs,rs2} in DIS,
and then to achieve the desired continuation to the saturation regime  
with the crucial help of lattice data\cite{ukqcd,rs-lat}. In a
complementary approach, we have considered the 
semi-classical, non-abelian eikonal approximation. It results in the
identification of the dipole with a Wilson loop, scattering
in the non-perturbative colour field of the proton. The field
$A_\mu^{(I)}\propto 1/\sqrt{\alpha_{\rm s}}$ due to background instantons was
studied as a concrete example, leading to 
results in qualitative agreement with the first approach. Due to the 
limitation of space, we focus on the first approach only and refer
to Refs.~\cite{su3,su4} for the second one.

\begin{figure}[t]
\parbox{4.1cm}{\includegraphics[width=4.1cm]{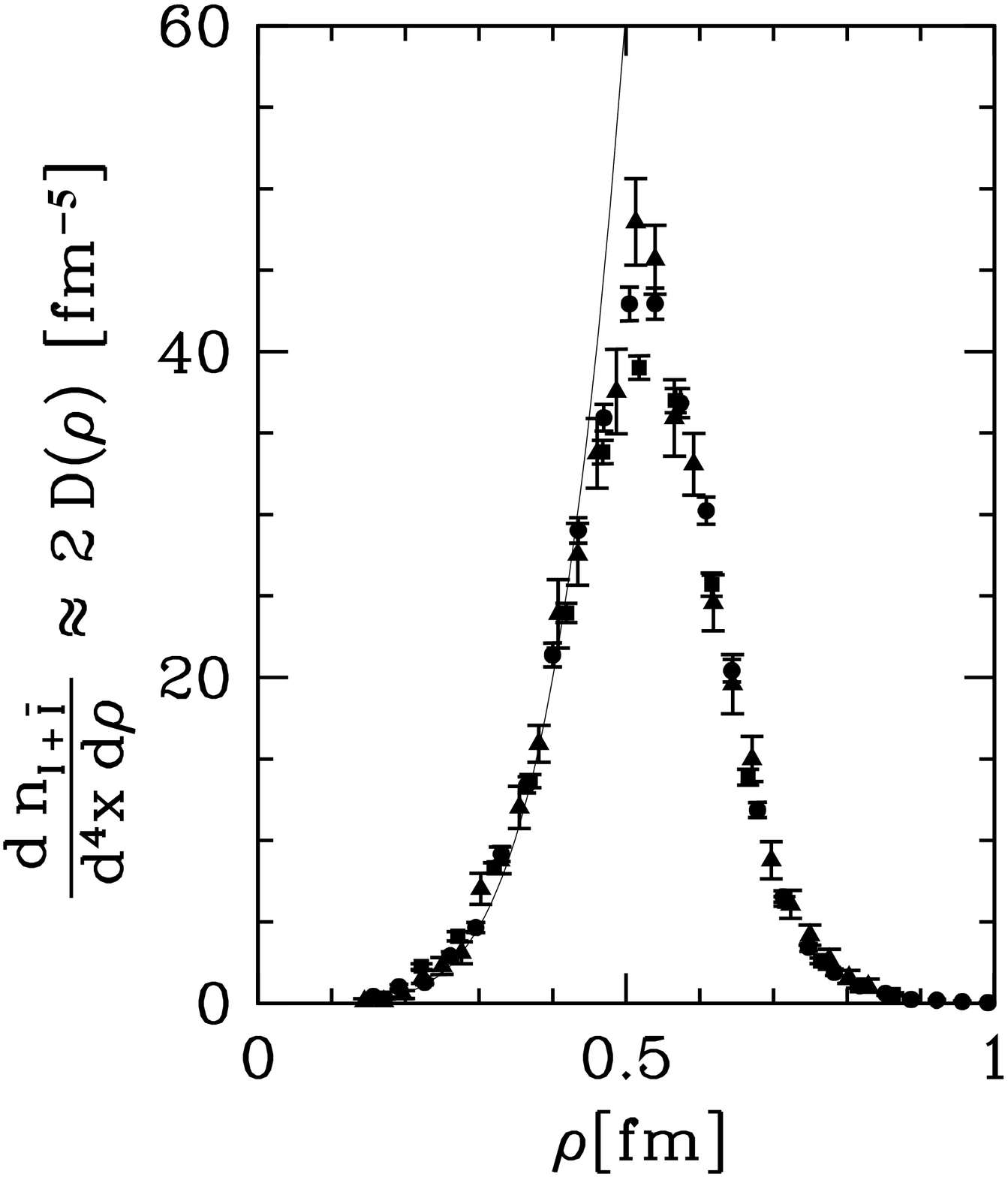}}   \hfill
\parbox{4.1cm}{\includegraphics[width=4.1cm]{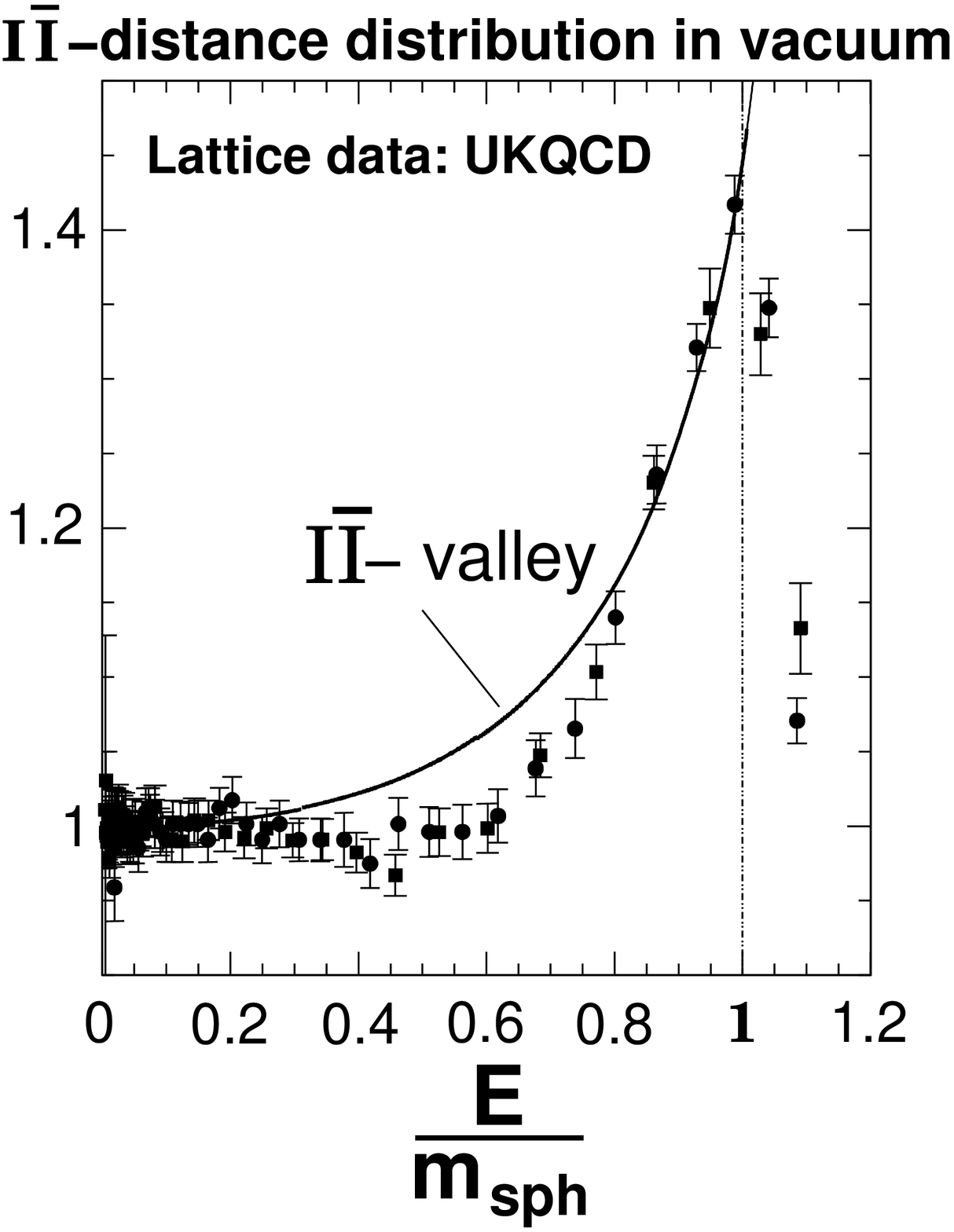}}\hfill   
\parbox{4.1cm}{\includegraphics[width=4.1cm]{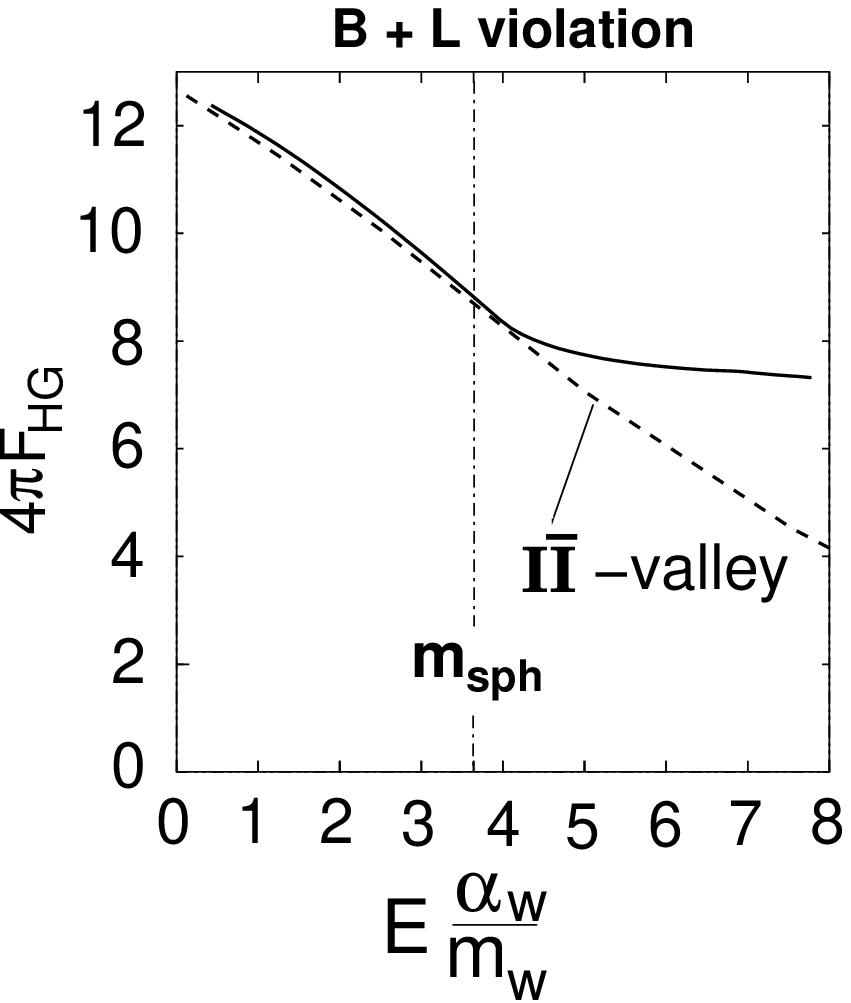}}
 \caption[dum]{(Left) UKQCD lattice data \cite{ukqcd,rs-lat,rs3} of
 the $(I+\bar{I})$-size distribution for quenched QCD ($n_f =
 0)$. Both the sharply defined $I$-size scale $\langle\rho\rangle
 \approx 0.5$ fm and the parameter-free agreement with
\mbox{$I$-perturbation} theory\cite{rs-lat,rs3} (solid line) for
$\rho\lwig 0.35$ fm are apparent.
(Middle) UKQCD lattice
 data~\cite{ukqcd,rs-lat}  of the (normalized) $I\bar{I}$-distance
 distribution and the corresponding $I\bar{I}$-valley
 prediction~\cite{su2} displayed versus energy in units of the QCD
 sphaleron mass $m_{\rm sph}$. 
 (Right) The same trend for electroweak $B +L$ 
  -violation is apparent from an independent
 numerical simulation of the suppression exponent for
	 two-particle collisions ('Holy Grail' function) $F_{\rm HG}(E)$
 ~\cite{rubakov,ringwald}    
 \label{pic2}}  
\end{figure}

Let us first consider briefly the simplest (idealized) $I$-induced process, 
\mbox{$\gamma^\ast\,g\Rightarrow q_{\rm R}\overline{q}_{\rm R}$}, with one
flavour only and no final-state gluons. More details
may be found in Ref.~\cite{su2}. Already this simplest case
illustrates transparently that in the presence of a background
instanton, the dipole cross section indeed saturates with a 
saturation scale of the order of the average $I$-size $\rav$.  
We start by recalling the results for the total $\gamma^\ast N$ cross
section within $I$-perturbation theory from Ref.~\cite{mrs},  
\vspace{-0.2cm}
\begin{eqnarray}
\sigma_{L,T}(\xbj,Q^2)&=&
\int\limits^1_{\xbj} \frac{d x}{x}\left(\frac{\xbj}{
x}\right)G\left(\frac{\xbj}{x},\mu^2\right)\int d \, t\; \frac{d\,
\hat{\sigma}_{L,T}^{\gamma^* g}(x,t,Q^2)}{d\, t};\,\label{general}\\[1ex] 
\frac{d\hat{\sigma}_{L}^{\gamma^* g}}{d  t}&=&\frac{\pi^7}{2}
\frac{e_q^2}{Q^2}\frac{\alpha_{\rm em}}{\alpha_{\rm
s}}\left[x(1-x)\sqrt{t u}\,  \frac{\mathcal{R}(\sqrt{-
t})-\mathcal{R}(Q)}{t+Q^2}-(t\leftrightarrow  u)\right]^{\,2}, \label{mrs}
\end{eqnarray}
with a similar expression for $d\,\hat{\sigma}_{T}^{\gamma^\ast\,g}/d\,t$. 
Here, $G\left(\xbj,\mu^2\right)$ denotes the gluon density and $L,T$
refers to longitudinal and transverse photons, respectively.
Note that Eqs.~(\ref{general}),~(\ref{mrs}) involve the ``master'' integral
$\mathcal{R}(\mathcal{Q})$ with dimension of a length,  
\vspace{-0.2cm}
\begin{equation}
\mathcal{R}(\mathcal{Q})=\int_0^{\infty} d\rho\;D(\rho)\rho^5(\mathcal{Q}\rho)\mbox{K}_1(\mathcal{Q}\rho).
\label{masterI}
\end{equation}
In usual $I$-perturbation theory, the $\rho$-dependence of the
$I$-size distribution $D(\rho)$ in Eq.(\ref{masterI}) is
known~\cite{th} for sufficiently small $\rho$, $D(\rho)\approx
D_{I-{\rm pert}}(\rho) \propto \rho^{6-\frac{2}{3}n_f},$ the power law
increase of which, leads to (unphysical) IR-divergencies from
large-size instantons. However, for sufficiently large
virtualities $\mathcal{Q}$ in DIS , the crucial factor
$(\mathcal{Q}\rho)\,K_1(\mathcal{Q}\rho)\sim e^{-\mathcal{Q}\rho}$
exponentially suppresses large size instantons and
\mbox{$I$-perturbation} theory holds, as shown first in
Ref.~\cite{mrs}. Replacing the $I$-size distribution in
Eq.~(\ref{masterI}) with the one from the lattice $D_{\rm
lattice}(\rho)$ (see Fig.~\ref{pic2} (left)), this restriction will
not be necessary anymore, whence $\mathcal{R}(0)=\int_0^{\infty}
d\rho\;D_{\rm lattice}(\rho)\rho^5\approx 0.3 \mbox{\ fm}$ becomes
finite and a $\mathcal{Q}^2$ cut is no longer required.

By means of a change of variables and a subsequent $2d$-Fourier
transformation, Eqs.~(\ref{general}), (\ref{mrs}) may indeed be
cast~\cite{su2} into a colour-dipole form (\ref{dipole-cross}). We obtain
e.g. for the longitudinal case, using $\hat{Q}=\sqrt{z\,(1-z)}\,Q$,
\begin{eqnarray}\
\lefteqn{\left(\left|\Psi_L\right|^2\sigma_{\mbox{\tiny DP}}\right)^{(I)}
 \approx\, \mid\Psi_L^{\rm pQCD}(z,r)\mid^{\,2}\,
\frac{1}{\alpha_{\rm s}}\,\xbj\,
G(\xbj,\mu^2)\,\frac{\pi^8}{12}}\label{resultL}\\[1ex] 
&&\times\left\{\int_0^\infty\,d\rho
D(\rho)\,\rho^5\,\left(\frac{-\frac{d}{dr^2}\left(2 r^2 
\frac{\mbox{K}_1(\hat{Q}\sqrt{r^2+\rho^2/z})}{\hat{Q}\sqrt{r^2+\rho^2/z}}
\right)}{{\rm K}_0(\hat{Q}r)}-(z\leftrightarrow 1-z)
\right)\right\}^2.\nonumber 
\end{eqnarray} 
The strong peaking of $D_{\rm
lattice}(\rho)$ around \mbox{$\rho\approx\rav$}, then implies 
\begin{equation}
\left(|\Psi_{L,T}|^{\,2}
\sigma_{\mbox{\tiny DP}}\right)^{(I)}\Rightarrow\left\{\begin{array}{llcl} 
\mathcal{O}(1) \mbox{\rm \ but exponentially small};&r\rightarrow 0\,,\\[2ex]
\mid\Psi^{\rm \,pQCD}_{L,T}\mid^{\,2}\,\frac{1}{\alpha_{\rm
s}}\,\xbj\,G(\xbj,\mu^2)\,\frac{\pi^8}{12}\,\mathcal{R}(0)^2;
&\frac{r}{\rav}\gwig 1\,	.\label{final} 
\end{array}\right.
\end{equation} 
Hence, the association of the intrinsic instanton scale $\rav$ with
the saturation scale $r_s$ becomes apparent from Eqs.~(\ref{resultL}),
(\ref{final}): $\sigma_{\mbox{\tiny DP}}^{(I)}(r,\ldots)$ rises strongly as function of
$r$ around $r_s\approx\langle\rho\rangle$, and  
indeed {\em saturates} for $r/\rav>1$  towards a {\em constant
geometrical limit}, proportional to the area
$\pi\,\mathcal{R}(0)^2\, =\,  \pi\left(\int_0^\infty\,d\rho\,D_{\rm
lattice}(\rho)\,\rho^5\right)^2$, subtended by the instanton.
Since $\mathcal{R}(0)$ would be divergent within
$I$-perturbation theory, the information about $D(\rho)$ from the 
lattice is crucial for the finiteness of the result. 

Now we turn to the more realistic process with an arbitrary number of
gluons and flavours in the final state, which causes a significant
complication. On the other hand, it is due to the inclusion of
final-state gluons that the identification of the QCD-sphaleron state
with the colour glass condensate has emerged~\cite{su2,su3}, with the
qualitative ``saturation'' features remaining unchanged.  Most of the
$I$-dynamics resides in the $I$-induced $q^\ast(q^\prime)\,
g(p)$-subprocesses with an incoming off-mass-shell quark $q^\ast$
originating from photon dissociation. The important kinematical
variables are the $I$-subprocess energy $E=\sqrt{(q^\prime+p)^2}$ and
the quark virtuality $Q^{\prime\,2}=-q^{\prime\,2}$.

It is convenient to account for the final-state
gluons by means of the so-called ``$I\bar{I}$-valley
method''~\cite{yung}. It allows to achieve via the optical theorem
 an elegant summation over the 
gluons in form of an exponentiation, with the  
effect of the gluons residing entirely in the $I\bar{I}$-valley interaction
\mbox{$-1\le\Omega_\text{valley}^{I\bar{I}}(\frac{R^2}{\rho\bar{\rho}}+\frac{\rho}{\bar{\rho}}+\frac{\bar{\rho}}{\rho};U)
\le 0\,.$} The new collective coordinate
$R_\mu$ denotes the $I\bar{I}$-distance, while the matrix $U$ characterizes the
relative $I\bar{I}$-colour orientation. Most
importantly, the functional form of $\Omega_\text{valley}^{I\bar{I}}$
is analytically known~\cite{kr,verbaarschot} (formally) for {\em all} values of
$R^2/(\rho\bar{\rho})$.

%%%%%%%%%%%%%%%%%%%%%%%%%%%%%%figure%%%%%%%%%%%%%%%%%%%%%%%%%%%%%%%
%%%%%%%%%%%%%%%%%%%%%%%%%%%%%%%%%%%%%%%%%%%%%%%%%%%%%%%%%%%%%%%%%%%
Our strategy here is identical to the one for the
``simplest process'': Starting point is the $\gamma^\ast N$ cross section, this
time obtained by means of the $I\bar{I}$-valley method~\cite{rs2}. The
next step is a variable and Fourier transformation into the
colour-dipole picture.
The dipole cross section $\tilde{\sigma}^{(I),{\rm
    gluons}}_{\mbox{\tiny DP}}(\vl^2,\xbj,\ldots)$ before the final
2d-Fourier transformation $\vl\leftrightarrow\vr$ to the dipole size
$\vr$, arises simply as an energy integral over the $I$-induced total
$q^\ast g$ cross section from Ref.~\cite{rs2},
\begin{equation}
\tilde{\sigma}^{(I),{\rm gluons}}_{\mbox{\tiny DP}} \approx
\frac{\xbj}{2}\,G(\xbj,\mu^2)\,\int_0^{E_{\rm max}}
\frac{d\,E}{E} \left[\frac{E^4}{(E^2+Q^{\,\prime 2})\,Q^{\,\prime 2}}\, 
\sigma^{(I)}_{q^\ast \,g}\left(E,\vl^2,\ldots\right)\right],
\label{sigdipglue}
\end{equation}
involving in turn integrations over the $I\bar{I}$-collective coordinates
$\rho,\bar{\rho},U$ and $R_\mu$.

In the softer regime of interest for saturation, we again substitute
$D(\rho) = D_{\rm lattice}(\rho)$, which enforces $\rho\approx\bar{\rho}\approx
\langle\rho\rangle$ in the respective $\rho,\bar{\rho}$-integrals,
while the integral over the $I\bar{I}$-distance $R$ is dominated by a
{\it saddle point}, 
\begin{equation}
\frac{R}{\rav} \approx {\rm
function}\left(\frac{E}{m_{\rm sph}}\right); \hspace{2ex} m_{\rm
sph}\approx \frac{3\pi}{4}\frac{1}{\alpha_{\rm s}\,\rav} =\mathcal{O}({\rm
\,few\ GeV\,}).
\label{sphaleron1}
\end{equation}
At this point, the mass $m_{\rm sph}$ of the
QCD-sphaleron~\cite{rs1,diak-petrov}, i.e the barrier height
separating neighboring topologically inequivalent vacua, enters as the
scale for the energy $E$. The saddle-point dominance implies a one-to-one relation, 
\begin{equation} 
\frac{R}{\rav} \Leftrightarrow \frac{E}{m_{\rm sph}}; \hspace{2ex}
\mbox{\rm with}\ R=\rav \Leftrightarrow E\approx m_{\rm sph}.
\label{sphaleron2}
\end{equation}
Our continuation to the saturation regime now involves crucial lattice
information about $\Omega^{I\bar{I}}$. The relevant lattice observable
is the distribution of the $I\bar{I}$-distance~\cite{rs-lat,su2} $R$,
providing information on $\left\langle\exp
[-\frac{4\pi}{\alpha_{\rm s}}\Omega^{I\bar{I}}]\right\rangle_{U,\rho,\bar{\rho}}$
in Euclidean space. Due to the crucial saddle-point
relation~(\ref{sphaleron1}),~(\ref{sphaleron2}), we may replace the
original variable $R/\rav$ by $E/m_{\rm sph}$. A comparison of the
respective $I\bar{I}$-valley predictions with the UKQCD lattice
data~\cite{ukqcd,rs-lat,su2} versus $E/m_{\rm sph}$ is displayed in
Fig.~\ref{pic2} (middle). 
It reveals the important result that the
$I\bar{I}$-valley approximation is quite reliable up to $E\approx
m_{\rm sph}$. Beyond this point a marked disagreement rapidly
develops: While the lattice data show a {\it sharp peak} at $E\approx
m_{\rm sph}$, the valley prediction continues to rise indefinitely for
$E\gwig m_{\rm sph}$! It is remarkable that an extensive recent and
completely independent semiclassical numerical
simulation~\cite{rubakov} shows precisely the same trend for
electroweak $B +L$-violation, as displayed in Fig.~\ref{pic2} (right).

It is again at hand to identify $\Omega^{I\bar{I}}=
\Omega^{I\bar{I}}_{\rm lattice}$ for $E\gwig m_{\rm sph}$.
Then the integral over the
$I$-subprocess energy spectrum  (\ref{sigdipglue}) in the dipole cross section appears to 
be dominated by the sphaleron configuration at $E\approx m_{\rm sph}$
The feature of saturation analogously to the ``simplest process'' in
Sec.~3.1 then implies the
announced identification of the colour glass
condensate with the QCD-sphaleron state.
%%%%%%%%%%%%%%%%%%%%%%%%%%%%%55


\begin{thebibliography}{99}
%\bibitem{HERA}
%C.~Adloff {\it et al.} [H1 Collaboration], 
%\Journal{\EPJC}{21}{33}{2001}; \\
%S.~Chekanov {\it et al.} [ZEUS Collaboration],
%\Journal{\EPJC}{21}{443}{2001}.
%\bibitem{DGLAP}
% V.N.~Gribov and L.N.~Lipatov,	
%\Journal{\SNP}{15}{438}{1972}; \\
% L.N.~Lipatov,
%\Journal{\SNP}{20}{94}{1975}; \\
%G.~Altarelli and G.~Parisi,
%\Journal{\NPB}{126}{298}{1977}; \\
%Y.L.~Dokshitzer,
%\Journal{\JETP}{46}{641}{1977}.
%\bibitem{BFKL}
%L.N.~Lipatov,
%\Journal{\SNP}{23}{338}{1976}; \\
%V.S.~Fadin, E.A.~Kuraev and L.N.~Lipatov,
%\Journal{\PLB}{60}{50}{1975},\\ 
%\Journal{\JETP}{44}{443}{1976}, 
%\Journal{\JETP}{45}{199}{1977}; \\
%I.I.~Balitsky and L.N.~Lipatov,
% \Journal{\SNP}{28}{822}{1978}.
\bibitem{gribovnu}
%        L.V.~Gribov, E.M.~Levin and M.G.~Ryskin, \Journal{\NPB}{188}{555}{1981}; \\
 L.V.~Gribov, E.M.~Levin and M.G.~Ryskin, \Journal{\PRP}{100}{1}{1983}.
\bibitem{cgc}
E.~Iancu, A.~Leonidov and L.~D.~McLerran,
\Journal{\NPA}{692}{583}{2001}.
%%CITATION = HEP-PH 0011241;%%
%E.~Ferreiro {\it et al.}, %E.~Iancu, A.~Leonidov and L.~McLerran,
%\Journal{\NPA}{703}{489}{2002}.
\bibitem{bpst}
A.~Belavin {\it et al.}, \Journal{\PLB}{59}{85}{1975}.
%%CITATION = PHLTA,B59,85;%%
\bibitem{su2}
F.~Schrempp and A.~Utermann, \Journal{\PLB} {\bf 543}{197} {2002}.
%%CITATION = HEP-PH 0207300;%%
\bibitem{su3}
F.~Schrempp and A.~Utermann, 
%{\it Proc. Strong and Electroweak Matter 2002}, Heidelberg, Oct. 2002,
%ed. M.G. Schmidt, p. 477 [
arXiv:hep-ph/0301177; arXiv:hep-ph/0401137.
%%CITATION = HEP-PH 0301177;%%
\bibitem{rs1}
A.~Ringwald and F.~Schrempp,
%{\it Proc. Q	uarks '94}, ed  D.Yu.~Grigoriev  {\it et al.} (Singapore:
%World Scientific) p 170, 
arXiv:hep-ph/9411217. 
\bibitem{mrs} S.~Moch, A.~Ringwald and F.~Schrempp, 
\Journal{\NPB}{507}{134}{1997}.
\bibitem{rs2} A.~Ringwald and F.~Schrempp, 
\Journal{\PLB}{438}{217}{1998}.
%%CITATION = HEP-PH 9806528;%% 
%, [arXiv:hep-ph/9806528].
\bibitem{qcdins} A.~Ringwald and F.~Schrempp, 
\Journal{\CPC} {132} {267} {2000}.
%%CITATION = HEP-PH 9911516;%% 
%, [arXiv:hep-ph/9911516].
%\bibitem{sphal2}
%H.~Aoyama and H.~Goldberg,
%\Journal{\PLB} {188}{506}{1987};
%A.~Ringwald, \Journal{\NPB} {330}{1}{1990}; O.~Espinosa,
%\Journal{\NPB} {343}{310}{1990}.
%%CITATION = PHLTA,B188,506;%%
%%CITATION = NUPHA,B343,310;%%
%%CITATION = NUPHA,B330,1;%%
%\bibitem{shuryak2}
%D.M.~Ostrovsky {\em et al.}, 
%\Journal{\PRD}{66}{036004}{2002}.
%%CITATION = HEP-PH 0204224;%%
%\bibitem{sphal1}
%F.~R.~Klinkhamer and N.~S.~Manton,
%\Journal{\PRD} {30}{2212}{1984}.
%%CITATION = PHRVA,D30,2212;%%
\bibitem{ukqcd}
D.A.~Smith and M.J.~Teper (UKQCD), \Journal{\PRD}{58}{014505}{1998}.
\bibitem{rs-lat} A.~Ringwald and F.~Schrempp, 
\Journal{\PLB}{459}{249}{1999}.
%%CITATION = HEP-LAT 9903039;%%
%, [arXiv:hep-lat/9903039].
\bibitem{rs3}
A.~Ringwald and F.~Schrempp,
\Journal{\PLB}{503}{331}{2001}.
%%CITATION = HEP-PH 0012241;%%
%, [arXiv:hep-ph/0012241].
\bibitem{levin}
D.E.~Kharzeev, Y.V.~Kovchegov and E.~Levin, \Journal{\NPA}{690}{621}{2001}.
\bibitem{shuryak1}
E.~Shuryak and I.~Zahed, \Journal{\PRD}{62}{085014}{2000}; \\
M.~Nowak, E.~Shuryak and I.~Zahed,
\Journal{\PRD}{64}{034008}{2001}. 
%%CITATION = HEP-PH 0012232;%%
\bibitem{shuryak3}
E.~Shuryak and I.~Zahed,
%``Understanding the non-perturbative deep-inelastic scattering:
%Instanton-induced inelastic dipole dipole cross section,''
Phys.\ Rev.\ D {\bf 69} (2004) 014011.
%%CITATION = HEP-PH 0307103;%%
%%CITATION = HEP-PH 0307103;%%
\bibitem{su1}
F.~Schrempp and A.~Utermann, \Journal{\APPB}{\bf 33}{3633} {2002}.
%%CITATION = HEP-PH 0207052;%%.
\bibitem{fs}
F.~Schrempp,
\Journal{\JPG}{\bf 28}{915}{2002}.
%%CITATION = HEP-PH 0109032;%%
\bibitem{dipole}
N.~Nikolaev  and B.G.~Zakharov,  \Journal{\ZPC}{49}{607}{1990};
\Journal{\ZPC}{53}{331}{1992}; A.H. Mueller,  
\Journal{\NPB}{415}{373}{1994}. 
\bibitem{dipole-pqcd}
F.~E.~Low, \Journal{\PRD}{12}{163}{1975}.
%%CITATION = PHRVA,D12,163;%%
\bibitem{su4}
F.~Schrempp and A.~Utermann, in preparation.
\bibitem{th}
G.~`t~Hooft, 
\Journal{\PRD}{14}{3432}{1976}.
%; \Journal{\PRD}{18}{2199}{1978} (Erratum).
\bibitem{yung}
A. Yung,
\Journal{\NPB}{297}{47}{1988}.
\bibitem{kr}
V.V. Khoze and A. Ringwald, 
\Journal{\PLB}{259}{106}{1991}.
\bibitem{verbaarschot}
J. Verbaarschot, 
\Journal{\NPB}{362}{33}{1991}.
\bibitem{diak-petrov}
D.~Diakonov and V.~Petrov,
\Journal{\PRD}{50}{266}{1994}. 
\bibitem{rubakov}
F.~Bezrukov {\em et al.},
\Journal{\PRD}{68}{036005}{2003}.
%%CITATION = HEP-PH 0304180;%%
\bibitem{ringwald}
A.~Ringwald, \Journal{\PLB}{555}{227}{2003}.
%%CITATION = HEP-PH 0212099;%%
%%CITATION = HEP-PH 0302112;%%
\end{thebibliography}
\end{document}